%% file: neurips_2021.tex
\documentclass{article}

\usepackage[preprint, nonatbib]{neurips_2021}

\usepackage[utf8]{inputenc} 
\usepackage[T1]{fontenc}    
\usepackage[ruled,vlined]{algorithm2e}
\usepackage{hyperref}       
\usepackage{url}            
\usepackage{booktabs}       
\usepackage{amsfonts}       
\usepackage{nicefrac}       
\usepackage{microtype}      
\usepackage{xcolor}         
\usepackage{graphicx}
\usepackage{multicol}
\usepackage{capt-of}
\usepackage{biblatex}

\title{Experimentally testable whole brain manifolds that recapitulate behavior}

  \author{%
  Gerald M. Pao*\\
  Scripps Institution of Oceanography, University of California San Diego\\
  La Jolla, CA 92093 USA\\
  \texttt{geraldpao@gmail.com} \\
   \And
   Cameron Smith \\
   University of California San Diego \\
   La Jolla, CA 92993 USA \\
   \texttt{cos008@ucsd.edu} \\
   \And
   Joseph Park \\
   CTBTO Preparatory Commission \\
   1400 Vienna, Austria\\
   \texttt{JosephPark@IEEE.org} \\
   \And
   Keichi Takahashi \\
   Nara Institute of Science and Technology \\
   Nara, Japan \\
   \texttt{keichi@is.naist.jp} \\
   \And
   Wassapon Watanakeesuntorn \\
   Nara Institute of Science and Technology \\
   Nara, Japan \\
   \texttt{wassapon.watanakeesuntorn.wq0@is.naist.jp} \\
   \And
   Hiroaki Natsukawa \\
   Kyoto University \\
   Kyoto, Japan \\
   \texttt{natsukawa.hiroaki.3u@kyoto-u.ac.jp} \\
    \AND
   Sreekanth H Chalasani \\
   Salk Institute, MNL \\
   La Jolla, California, USA \\
   \texttt{schalasani@salk.edu} \\
    \And
   Tom Lorimer \\
   University of California San Diego \\
   La Jolla, California 92093 USA \\
   \texttt{tom.lorimer@gmail.com} \\
    \And
   Ryousei Takano \\
   Digital Architecture Research Center, AIST \\
   Tokyo, Japan \\
   \texttt{takano-ryousei@aist.go.jp} \\
    \And
   Nuttida Rungratsameetaweemana \\
   Salk Institute, CNL \\
   La Jolla California 92037, USA \\
   \texttt{nrungrat@salk.edu} \\
   \And
   George Sugihara \\
  Scripps Institution of Oceanography, University of California San Diego\\
   La Jolla, CA 92093, USA \\
   \texttt{gsugihara@gmail.com} \\
}

\begin{document}

\maketitle

\newpage

\begin{abstract}
Brain activity is a highly nonlinear phenomenon, and as such, dynamical relationships between neurons and expressed behaviors have been described on low dimensional attractor manifolds mostly through dimensionality reduction techniques. These range from simple principal component analysis to sophisticated sequential recurrent variational autoencoders such as LFADS. These low dimensional representations i.e. the principal components of PCA or factors of a variational autoencoder, are a dimensionally reduced representation of the original high dimensional activity of thousands of observed neurons. Although these low dimensional manifolds offer unifying views of global brain activity the principal components of PCA reduced or factors of variational autoencoders are not experimentally tractable. For example with PCA, it is not possible to experimentally manipulate the factional amount of the activity of a particular neuron(e.g. 0.7 of neuron1, 0.9 of neuron 2, 0.3 of neuron 3 etc…) that corresponds to its projection on a principal component. Thus although these low dimensional manifolds provide appealing explanatory power, it is ultimately difficult if not impossible to address these experimentally. To solve this problem, here, we propose an algorithm grounded in dynamical systems theory that generalizes manifold learning from a global state representation, to a network of local interacting manifolds – termed a Generative Manifold Network (GMN). Manifolds are discovered using the convergent cross mapping (CCM) causal inference algorithm which are then compressed into a reduced redundancy network. The representation is a network of manifolds embedded from observational data where each orthogonal axis of a local manifold is an embedding of a individually identifiable neuron or brain area that has exact correspondence in the real world. As such these can be experimentally manipulated to test hypotheses derived from theory and data analysis. Here we demonstrate that this representation preserves the essential features of the brain of a fly. In addition to accurate near-term prediction, the GMN model can be used to synthesize realistic time series of whole brain neuronal activity and locomotion viewed over the long term. Thus, as a final validation of how well GMN captures essential dynamic information, we show that the artificially generated time series can be used as a training set to predict out-of-sample observed fly locomotion, as well as brain activity in out of sample withheld data not used in model building. Remarkably, the artificially generated time series show realistic novel behaviors that \textit{do not} exist in the training data, but that \textit{do} exist in the out-of-sample observational data. This suggests that GMN captures inherently emergent properties of the network. We suggest our approach may be a generic recipe for mapping time series observations of any complex nonlinear network into a model that is able to generate naturalistic system behaviors that identifies variables that have real world correspondence and can be experimentally manipulated.

\end{abstract}

\section{Introduction}
\label{intro}

Neuroscience has a long history of studying oscillations \cite{Draguhn2004} and intermittent activity that may be compatible with descriptions of neural activity as dynamics on an attractor. In recent years an accumulating body of evidence shows that a large number of neuronal activity dynamics are well described in terms of low dimensional attractor manifolds \cite{Churchland2012,Low2018,Maheswaranathan2019,Brennan2019,Chaudhuri2019,Tajima2015}. Most of the literature finds these low dimensional manifolds within the ensemble properties of relatively large populations of neurons. Through dimensionality reduction approaches, most notably principal component analysis, one commonly finds that the low dimensional embedding contains surprisingly coherent structure \cite{Churchland2012,Bruno2017}. It is in these low dimensional embeddings derived from larger populations, where most observations have been made that show remarkable coherence when projected into low-dimensional space on the surfaces of manifolds. However these low dimensional representations where the factors or components are linear or nonlinear ensemble combinations of fractional activities of hundreds or thousands of neurons do not allow for experimental testing of the reduced dimensionality manifold models. 
We develop an algorithmic data-driven approach, based on networks of manifolds, that transforms empirical observations of neuronal activity into a realistic generative model where individual components are identifiable and correspond to real brain areas or neurons that can be manipulated experimentally.  

We demonstrate and validate our approach here on publicly available data, consisting of simultaneous time series recordings of near-whole-brain neural activity and movement (walking) in transgenic \textit{Drosophila melanogaster} from Ref. \cite{Aimon2019}. From these data we create a model that can generate walking behaviors in the plane that are not only similar to those observed in the real fly, but that also contain realistic observed fly movement behaviour that was not used in model training. 
Among several strong validations, our generative model most surprisingly produces artificial generated time series that exhibit realistic fly resting state behaviors that were not present in the traning data. 
This suggests that this type of network may embody emergent properties hidden in the brain that are not explicitly represented in the training set used to build the model. 

\section{Background and related work}
\label{background}

\subsection{Manifolds in brains}
The presence of coherent dynamics in neural activity has been recognized for at least two decades in one form of another, however much excitement has been generated in the last few years upon discovering how ubiquitous these dynamics are. 
\cite{Churchland2012,Tajima2015,Bruno2017}. These early manifolds, however appealing, were observations of coherent behavior without explicit dynamic information. More recently, the view of attractor dynamics on manifold surfaces has started to deliver predictive power. 
\cite{Low2018}. Variability in neural activity, that was commonly thought to be noise, was shown to be largely accounted for by the trajectory deviations due the existence of additional dimensions on the surface of the manifold when compared to the arena in which the rats were moving.Most methods allow for short term prediction and use dimensionality reduction to have a single latent low dimensional manifold that abstracts global population behavior of many neurons. Most remarkable is LFADS \cite{pandarinath2018inferring} which using a variational autoencoder not only predicts neural activity but also generates realistic neural dynamics using the factors at its bottleneck which lie on a low dimensional manifold. However these factors, like PCA components, or Fourier modes are abstractions that have no direct correspondence with real brain structures or neurons that allow experimental manipulation of the latent manifold. To enable experimental testing of these mathematical objects we developed the GMN framework that has a 1:1 correspondence of observations and variables that determine dimensions of local manifolds. A measure of the success and completeness of the model will be its ability to reproduce its output as well as its relevant internal dynamics.

\section{Embedding whole brain activity}
\label{embeddingBrain}

\subsection{Data}
To predict movements of an organism using whole brain activity requires a dataset that has simultaneous recordings of brain activity as well as the recorded motions of the organism. Currently given experimental limitations this is only possible with a few experimental species of animals. In our work we chose a dataset of from a headfixed Drosophila melanogaster on a Styrofoam ball where whole brain activity was recorded using light field microscopy of signals derived from a transgenic fluorescent calcium indicator GCaMP6f and walking motions were recorded from the rotation of the Styrofoam ball \cite{Aimon2019}. The whole brain dataset does not achieve single neuron resolution so the whole brain activity was segmented using independent component analysis to identify 80 different areas in the whole brain that were deemed to have distinct activities. These time series, together with the recorded left-right and forward speed of motion of the flies on the Styrofoam ball, comprise our data. At this stage, the data are split into training and testing sets (see methods in supplement).

\subsection{Algorithm}
To proceed from these individual time series to a network, we need to first understand how the time series relate to one another. To this end, we use convergent cross mapping (CCM) \cite{Sugihara2012}, a Takens theorem based causal inference method that can detect nonlinear causation between time series (even without correlation). We ran CCM on all pairs of time series in our \textit{training} data, producing for each pair a scalar value, $\rho_{\mbox{ccm}ij}$ that quantifies the nonlinear causal effect of time series $j$ on time series $i$. To extract from these causal effects the most salient, non-trivial and nonlinear, we subtracted from $\rho_{\mbox{ccm}}$ the absolute value of the linear (Pearson) correlation between each pair of (training) time series to generate a new matrix $\rho_{\mbox{diff}ij} = \rho_{\mbox{ccm}ij}-|\rho_{\mbox{corr}ij}|$. Thus $\rho_{\mbox{diff}}$ quantifies the nonlinear predictability skill over and above linear correlation, thereby identifying the most important causal links between the time series, while eliminating highly correlated time series from consideration containing redundant information. Conceptually one could think of it as a nonlinear analog of independent component analysis (ICA) where one finds non-redundant, nonlinear causal sources contributing to a given time series. This information is then used to combine timeseries into local manifolds, and to link these manifolds into a network though the inferred causal interactions between manifolds.

The manifold network is constructed around a given target variable $i$ (e.g. forward movement of the fly) as follows. First, the various time series $j$ with the highest values of $\rho_{\mbox{diff}ij}$ are identified. A number $E-1$ of these are selected, such that the final dimensionality of the manifold is $E$ (each time series will contribute one dimension). The selected variables together with the target variable can then define a manifold that will eventually be used to predict the target variable $i$ through attractor reconstruction using the Simplex method \cite{Sugihara1990a}. We move from this single manifold to a manifold network by noticing that each of the variables $j$ that is contributing to the manifold used to predict variable $i$, can in its turn also be predicted from another manifold that includes the variables that are most causally relevant to $j$, by the same scheme. Thus the manifold network is grown outwards from a target variable (e.g. forwards movement) by linking manifolds together via their prediction outputs.

The manifold, or attractor, of each node in our network is defined by the training data time series (``library'') for the variables used at that node. This time sequence of points in an $E$ dimensional space can be used to predict forwards in time from a novel point in this same $E$ dimensional space using a Simplex method . A brief example will clarify this prediction process. Suppose an attractor at a node that predicts left-right motion is defined using the training data for left-right motion, and 3 other variables. Given a new point in this 4 dimensional space (i.e. simultaneous values of left-right motion and the three other variables), we try to predict the values of those variables at the next time step. To do this, we find the closest 5 points on the training attractor (which define a 4 dimensional simplex), then iterate those training points forward by one time step. The predicted next position of the novel point is then center of mass of the forward-iterated simplex. The predicted value of left-right motion is thus simply the left-right motion component of this predicted vector. See Ref. \cite{Sugihara1990a} for details. When running the network in predictive mode (for model testing and validation) the novel point at each time step will come from testing set data. In generative mode, the novel point at each time step comes from the previous predicted output for each variable. To avoid instabilities when operating in generative mode, the novel point is nudged to its nearest neighbor on the training data manifold surface every $s$ time steps. In many cases observations might be incomplete and have missing observations to properly reconstruct a manifold of the correct dimensionality. The Takens theorem has a solution for this, as one can use time lagged time series of the observed variable as a substitute for the missing observation. In it's generalized form, the Takens theorem allows for the combination or real observations with lags of the observed variable time series\cite{Deyle2011}.Thus depending on the completeness of observations we used observations or a combination of observations and lags for attractor reconstruction.

At the network level, two details need to be addressed regarding data input and synchrony. Each node (or manifold or attractor) in the network is used to predict one target variable. This target variable is the only one fed directly into the node (either from the previous prediction in generative mode, or from a test dataset in prediction mode). The other variables corresponding to the node's attractor are fed in from its upstream nodes, which means that they need to be fed in one timestep ahead to maintain synchrony with the downstream node. For sets of nodes connected in a loop, this synchrony requirement can not be met, and so it is ignored, sending instead the predicted value at the current timestep to the downstream node. In the present implementation, nodes that are not reachable from the target node in a acyclic directed graph are not included in the network.(Figure 1)
\begin{figure}
  \centering
  \includegraphics[width=.8\columnwidth]{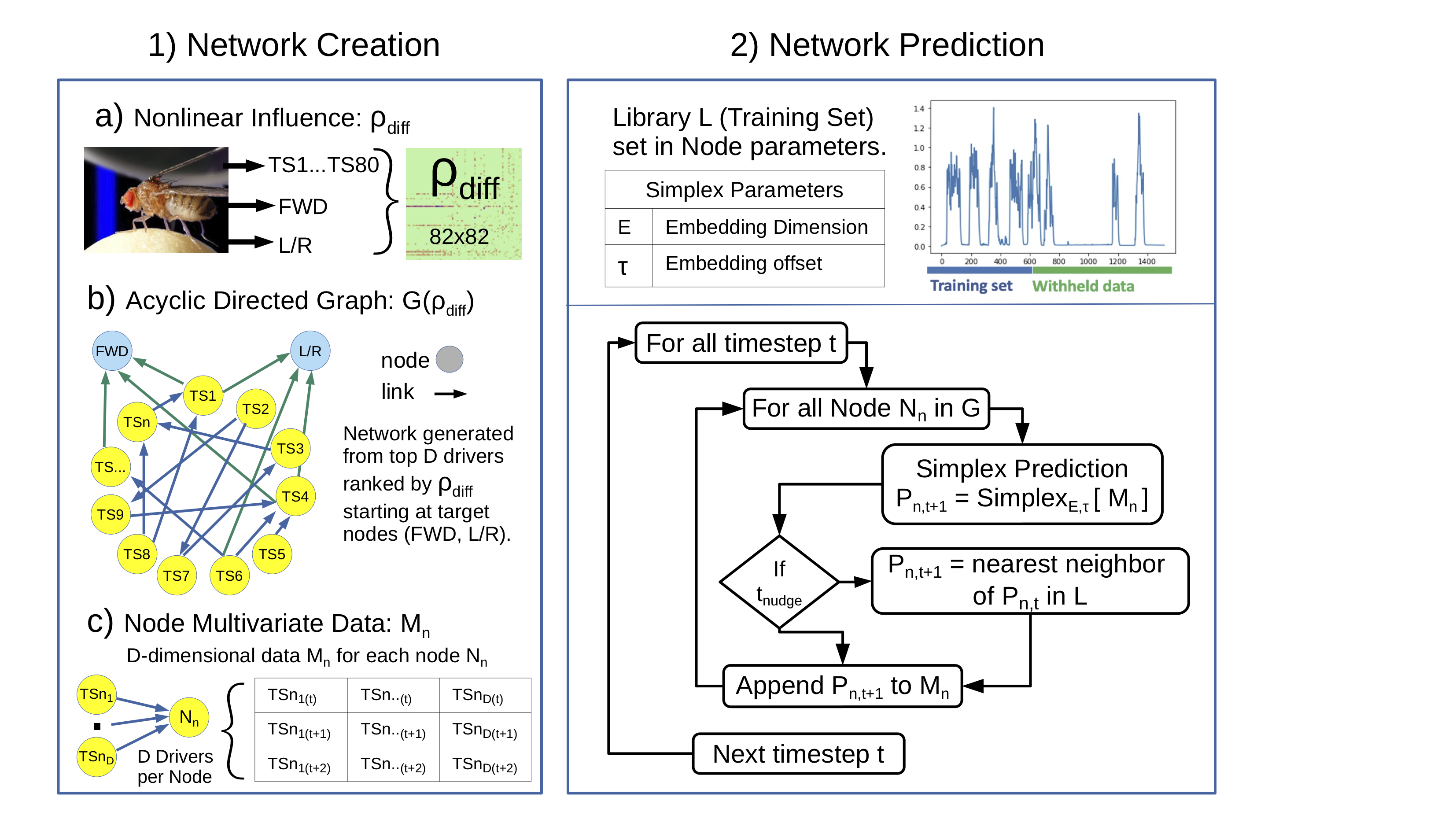}
  \caption{Schematic of the Generative Manifold Network (GMN) algorithm. 80 time series of neural activity, and 2 of the physical walking are used to  construct a $\rho_{\mbox{diff}}$ matrix as specified in the text. This selects for causation that cannot be explained by linear correlation, so that our network dynamics will not be dominated by synchrony. 1b) Each time series is identified with a node. Here we build a network to predict the FWD and L/R walking motion (blue nodes, arbitrary choice). Each blue node $i$ receives $D-1$ incoming connections from the yellow nodes $j$ that have the $D-1$ highest values of $\rho_{\mbox{diff}ij}$ . Then, each of the yellow nodes $j$ just connected receives $D-1$ incoming connections from the remaining not-connected yellow nodes $k$ that have the highest values of $\rho_{\mbox{diff}jk}$. This process repeats until all nodes are connected. 1c) The network structure defines the $D$-dimensional data $M_n$ from which that node’s manifold will be constructed. 2) The simplex prediction parameters $E$ and $\tau$ define the structure of the local training manifold: each point has $D$ variables each with $E$ time lags, with $\tau$ time steps per lag. The “Simplex” step in the prediction at each node follows the simplex of the $ED+1$ closest points to the last predicted point on the training manifold forward in time, and returns the center of mass of that forward-iterated simplex as a prediction.}
\end{figure}

\begin{algorithm}
\DontPrintSemicolon

\KwIn{Set of training time series $\{\mathbf{x}_{i}\}$}
\KwIn{Set of target node indices $tni = \{i_1,i_2,...\}$}
\KwIn{Number of drivers $D$}
\KwIn{Manifold embedding dimension $E$}
\KwIn{Number of points to generate $ng$}
\KwIn{Nudge interval $s$}
\KwOut{Set of generated time series $\{\mathbf{p}_{i}\}$}
\;
\KwData{Directed acyclic graph $G$}
\KwData{Queue of nodes $Q=tni$}
\;
Calculate pairwise CCM matrix $\rho_{\mathrm{ccm}_{ij}} = \mathrm{CCM}(\mathbf{x}_i, \mathbf{x}_j)$ \;
Calculate pairwise Pearson correlation matrix $\rho_{\mathrm{corr}_{ij}} = \mathrm{corr}(\mathbf{x}_i, \mathbf{x}_j)$\;
Causality matrix $\rho_{\mathrm{ccm}_{ij}} = \max(\rho_{\mathrm{diff}_{ij}}, 0) - | \rho_{\mathrm{ccm}_{ij}} |$\;
\;
\tcp{Network Construction}
\While{$Q \neq \emptyset$}{
    $i \leftarrow \mathrm{pop}(Q)$\;
    Set of top $D-1$ driving nodes $\{d_1, d_2, \dots d_{D-1}\}$ of $i$ corresponding to top $D-1$ values in $\rho_{\mathrm{ccm}_{ij}}$\;
    \For{driving node $d \in \{d_1, d_2, \dots d_{D-1}\}$ }{
        \If{$d \not\in G$ and edge $(d, i)$ does not introduce a cycle in $G$}{
            Add edge $(d, i)$ to $G$\;
            $\mathrm{push}(Q, d)$
        }
    }
}
\;
\tcp{Network Prediction}
Calculate topological ordering of G $to$\;
\For{$t \in [0, ng)$}{
    \For{node $i \in to $ }{
        $\mathbf{p}_{i, t+1} \leftarrow \mathrm{Simplex}_{E,\tau}(\mathbf{x}_i, \mathbf{x}_{d_1}, \mathbf{x}_{d_2}, \dots, \mathbf{x}_{d_D-1} )$\;
        \;
        \tcp{Nudging}
        \If{$t \bmod s = 0$}{
            $\mathbf{p}_{i, t+1} \leftarrow$ Nearest neighbor of $\mathbf{p}_{i, t+1}$ in $\mathbf{x}_i$\;
        }
    }
}
\caption{Generative Manifold Network algorithm}%
\label{alrc_algorithm}
\end{algorithm}

\section{Results}
\label{results}


\begin{figure}
  \centering
  \includegraphics[width=.75\columnwidth]{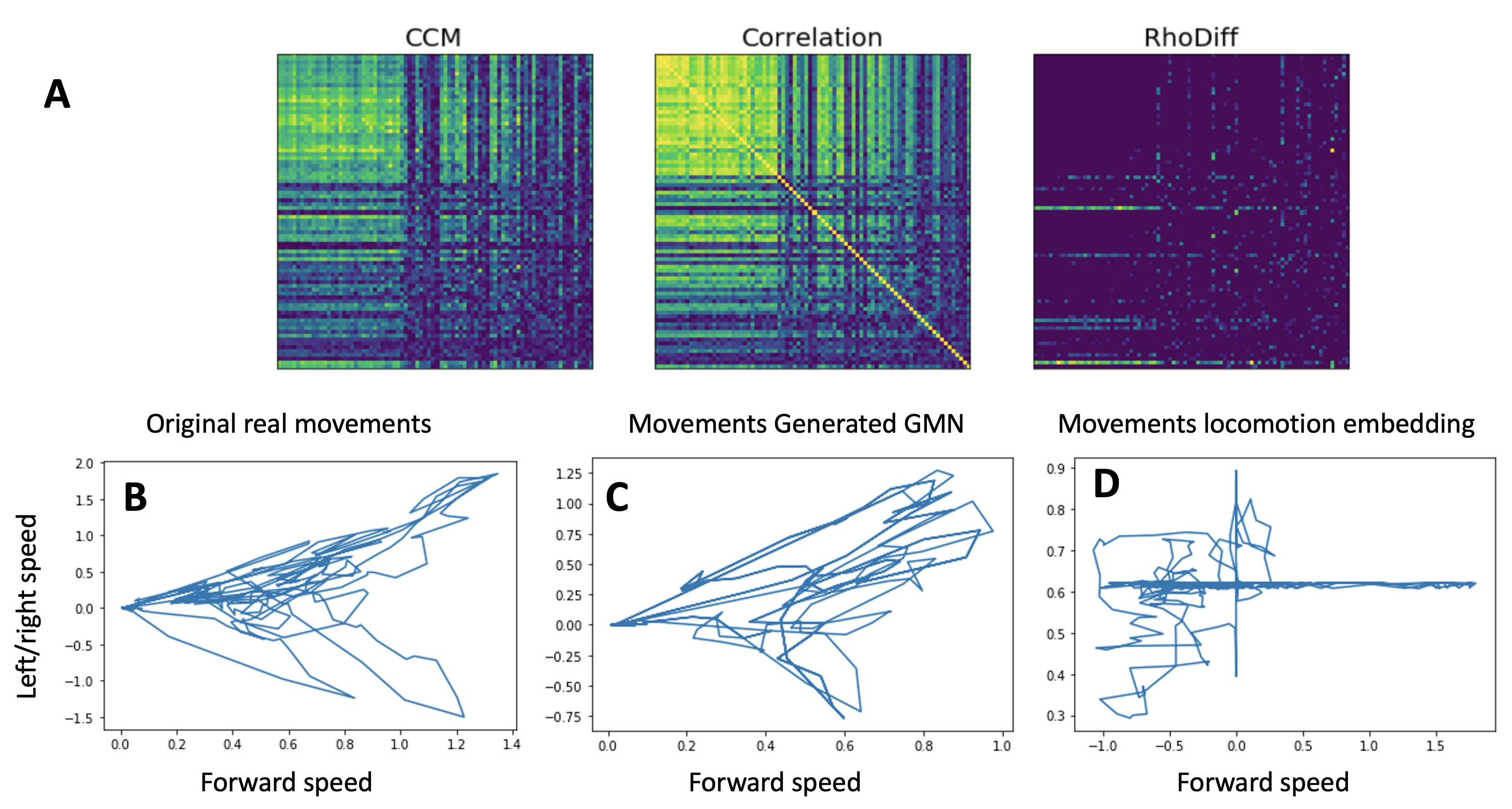}
  \caption{(A) To identify relationships between variables of a given manifold we obtained the CCM matrix and subtracted the absolute value of the correlation matrix to obtain the $\rho_{\mbox{diff}}$ residual. This identifies the candidate time series for each embedding. The phase portrait of the fly forward and left right speeds (B) are visually similar to the output of the generative manifold network (C). An embedding of prior movements of the fly fail to generate realistic fly speed changes and (D) suggests that the information comes from the brain and not from the walking patterns themselves.}
\end{figure}

\subsection{Generative properties}
Although Takens type time delay embeddings frequently have high predictive skill, they almost invariably fail over time if predictions are used recursively to drive the network forward (Figure 3C). Over time predictions decrease in amplitude and eventually converge to a constant, or to a very low amplitude oscillation. This is largely due to the fact that simplex projection is an averaging approach where projecting into a center of a simplex averages the result until there will be a projected point for which the nearest neighbors will never change. This is true for predicting itself as well as cross predicting to another variable as seen in Figure 3C, right panel, where the generated output converges to a constant. On the other hand time series outputs of the generative manifold network (Figure 3B,D) tended to not fall either into cycles nor converge to a constants and look superficially similar to the oscillations observed in the training set. The similarity of generated time series was not limited to movement regressors: also neural activity time series looked quite similar (Figure 3D). 
To have a more quantitative measure of the properties of the forward oscillations generated compared to the real forward speed time series we calculated the Fourier transform of the real time series and compared to the spectrum of the manifold network generated time series as well as the time series generated by the highest performing single brain area (Figure 4). Overall the manifold network generated time series data matched well where the cross correlation coefficient between the real spectrum and the manifold network generated spectrum was 0.92 whereas a time series from the brain area that best predicted forward movement was only able to obtain a correlation coefficient of 0.37.

\begin{figure}
  \centering
  \includegraphics[width=1\columnwidth]{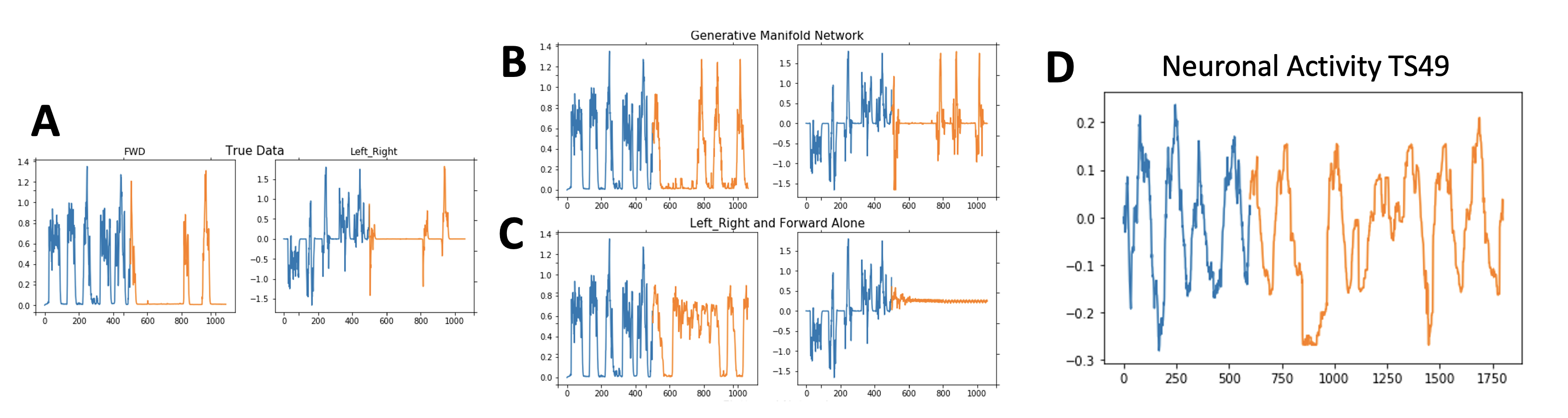}
  \caption{Recursively generating output (training interval in blue, testing interval in orange). (A) Real forward and left right speeds of Drosphila locomotion. (B) GMN generated speed time series. Note the emergence of pauses that are not in the training interval (see also Figures 5, 6). (C) Simplex generated time series from an embedding of the fly movements alone. (D) In addition to movements, the GMN can generate realistic neuronal activity.}
\end{figure}

\begin{figure}
  \centering
  \includegraphics[width=.75\columnwidth]{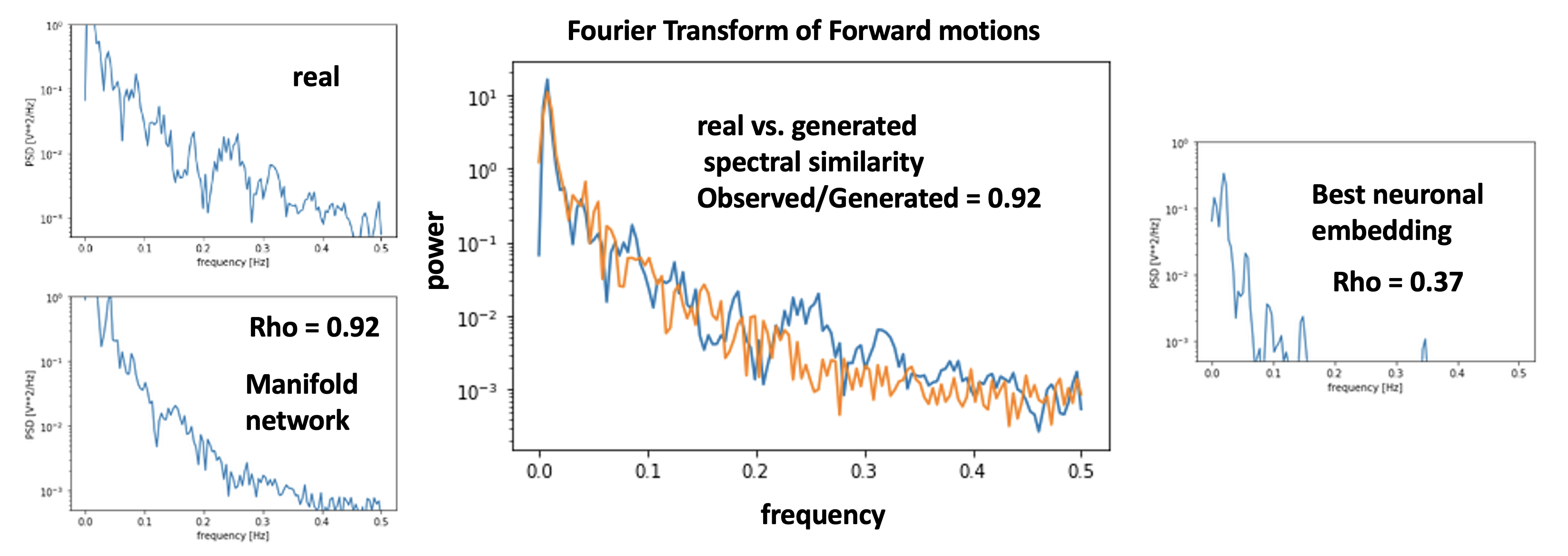}
  \caption{Comparison of the Fourier transform of the fly movement comparing real, generated and the best single neuronal predictor of forward walking speed.}
\end{figure}

\subsection{Movement information vs. brain information}
Here we compare the similarity of the forward and left right phase plots of the observed fly movement with the movements generated from the generative manifold network embedding. As seen in Figure 2, both the natural and artificial phase plots have a similar overall geometry and appear to be similar. To exclude the possibility that just the walking motion of the fly itself contains sufficient information to generate realistic fly movements, we embedded the left right and forward time series alone into a predictive manifold that ran in generative mode. This phase plot fails to give the same realistic geometry, suggesting that the information for realistic movement is contained in the neural activity patterns and not in the movements themselves. 

\begin{figure}
  \centering
  \includegraphics[width=1\columnwidth]{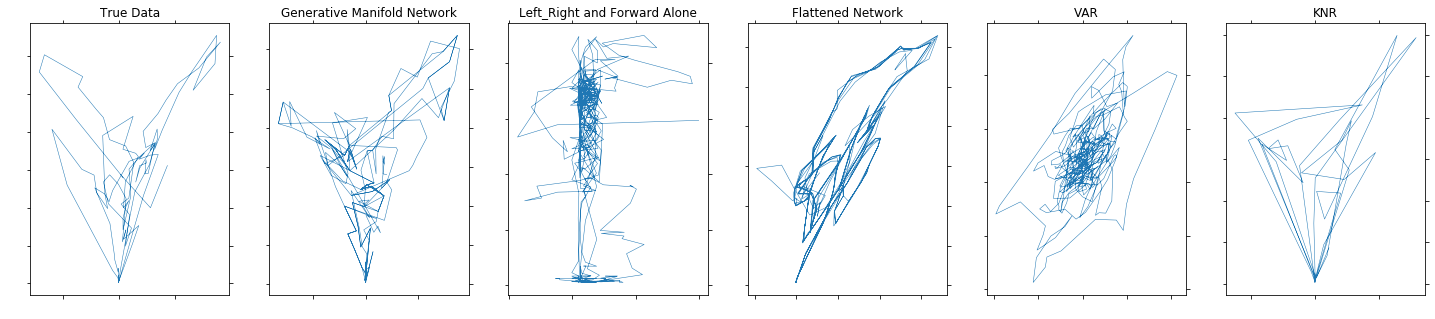}
  \caption{Comparison of the real fly speed phase portraits generated by: the GMN, an embedding of the fly left right and forward speed, the flattened network of 82 time series, a recursive prediction vector autoregressive model, and a recursive kNearest Neighbor predictor.}
\end{figure}

\subsection{Possibly emergent properties in the generative manifold network}
To build the generative manifold network model we used the first 500 points of the original downsampled time series out of 1200 total time points as seen in (Figures 6, 7b) left. The green underlined section shows the original time series and the orange underline are the network generated forward speed time series. Interestingly although training set did not have long movement pause periods (Figure 7b), the generated forward time series had a couple long low forward velocity down states. These low speed periods were also observed in the withheld real data that was not part of the training set. This suggests that the network contains the capability to generate behaviors that are not strictly present in the training set and that it is not simply reproducing behaviors in the training set. (Figure 7d) shows the generated time series for forward speed (Blue) and left right speed (red) for a stretch of generated data. As one can observe these rest periods actually experience some unnatural low amplitude oscillations. These oscillations might be masked in the real flies and might occur as subthreshold events in the real fly. Since none of these features appear in the original training set period, these observations suggest that this type of network is capable of generating emergent properties rather than just reproducing features of the training sets through features hidden in the network embeddings.

\subsection{Comparison of generated properties with other low dimensional embeddings}
In order to investigate the possible properties of the network we constructed, we compared the generative manifold network output to the real withheld data, the output of an embedding of the left right and forward speeds, and a flat network which is an embedding of all 80 time series of the entire dataset into a 80 dimensional embedding (Figure 5). Shown are the forward and left predictions for all 4 models (Figure 5). As can be seen both the left right and forward embeddings generated time series fall into cyclical behaviors from which they seem to be unable to get out, or converge to a constant value (Figures 5, 7). This does not appear to happen to the generative manifold network. Comparison of individual the left right and forward generated movements show that the generative manifold network appears to be the most realistic with the embeddings of the movements alone based on it’s past history being the most dissimilar. The fact that movements alone cannot generate realistic movements in a multidimensional embedding suggests that motions do not encode their own future movements and it is not Markov-chain like. In addition the fact that the generative manifold network performs better than the flat network which contains in principle all the information within the network suggests that the information gating within the network plays an important role.

\begin{figure}
  \centering
  \includegraphics[width=.7\columnwidth]{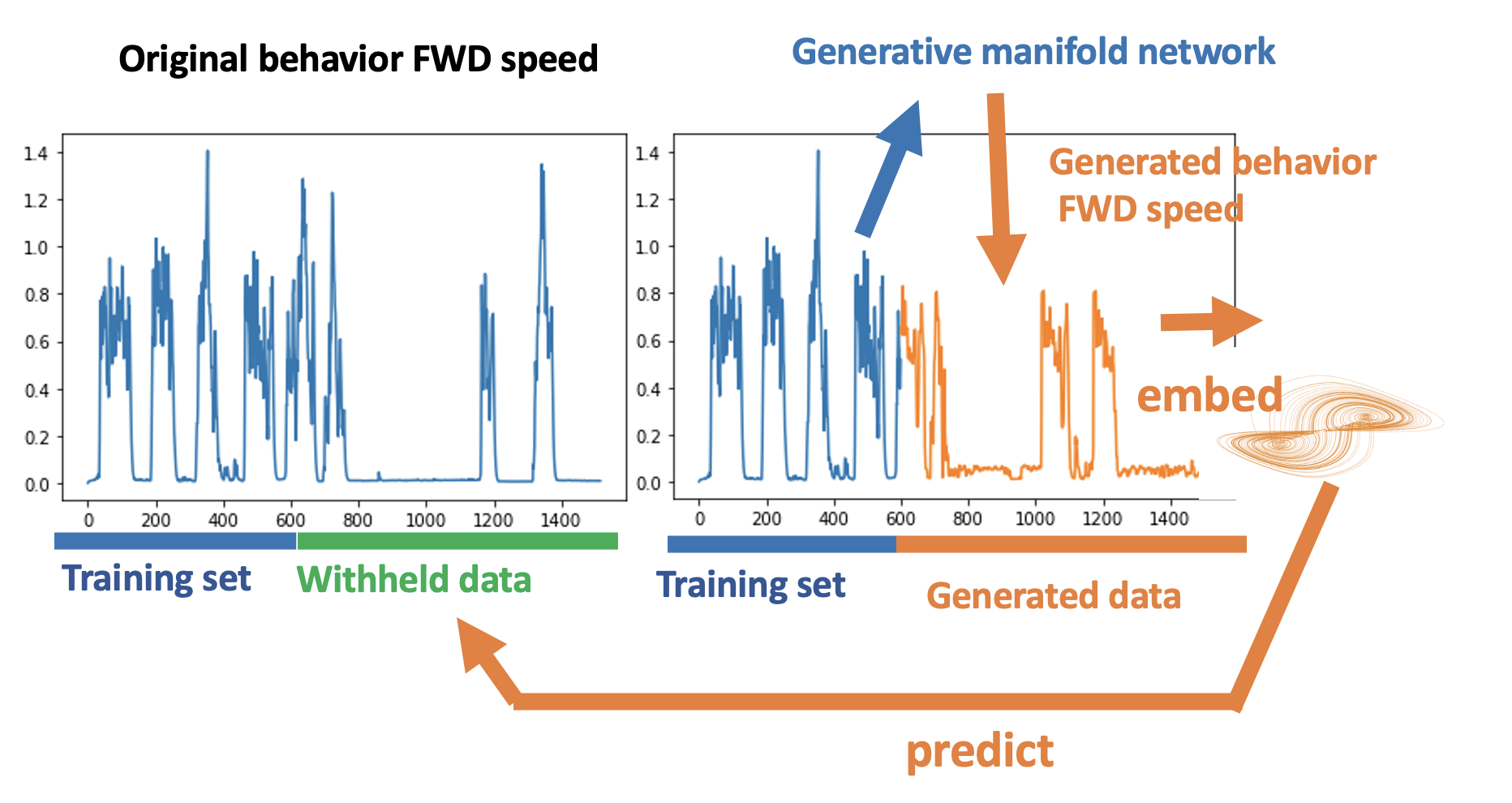}
  \caption{Validation scheme of the GMN generated data. The realism of the generated data is evaluated as it’s prediction skill to predict out of sample withheld time series.}
\end{figure}

\subsection{Generated time series can predict withheld real dynamics}
A stringent way to test that the time series that are generated by this network are realistic is to show that the artificially generated time series can be used as training data for the prediction of withheld real data of its own original time series: 
successful predictions indicate shared dynamics (Figure 6). For this test we generate time series data for every brain area as well as simultaneous time series of the forward and left right movements and compared the generative manifold network to: 1. an autoregressive linear model, 2. the flat network and 3. a 80 dimensional nearest neighbor predictor for the prediction of each brain area. Results in Figure 7 show that the generative manifold network outperforms all three: the flat network, the autoregressive linear model as well as the K-nearest neighbor model by significant margins indicating that GMN produces synthesized time series that have dynamical features that are closer to those in the the observed data than the alternatives (even more obvious if compared individually see suppl.data). Figure 5 shows a sample of the predictions from the GMN, the embedded left right and forward motion speeds of the fly and the flattened network which embeds all time series into a single manifold. The ability of the generated data to skillfully predict withheld real data from the same brain areas shows that the generated data is indeed quite similar to data directly obtained from the fly (Figures 5, 7).

\begin{figure}
  \centering
  \includegraphics[width=1\columnwidth]{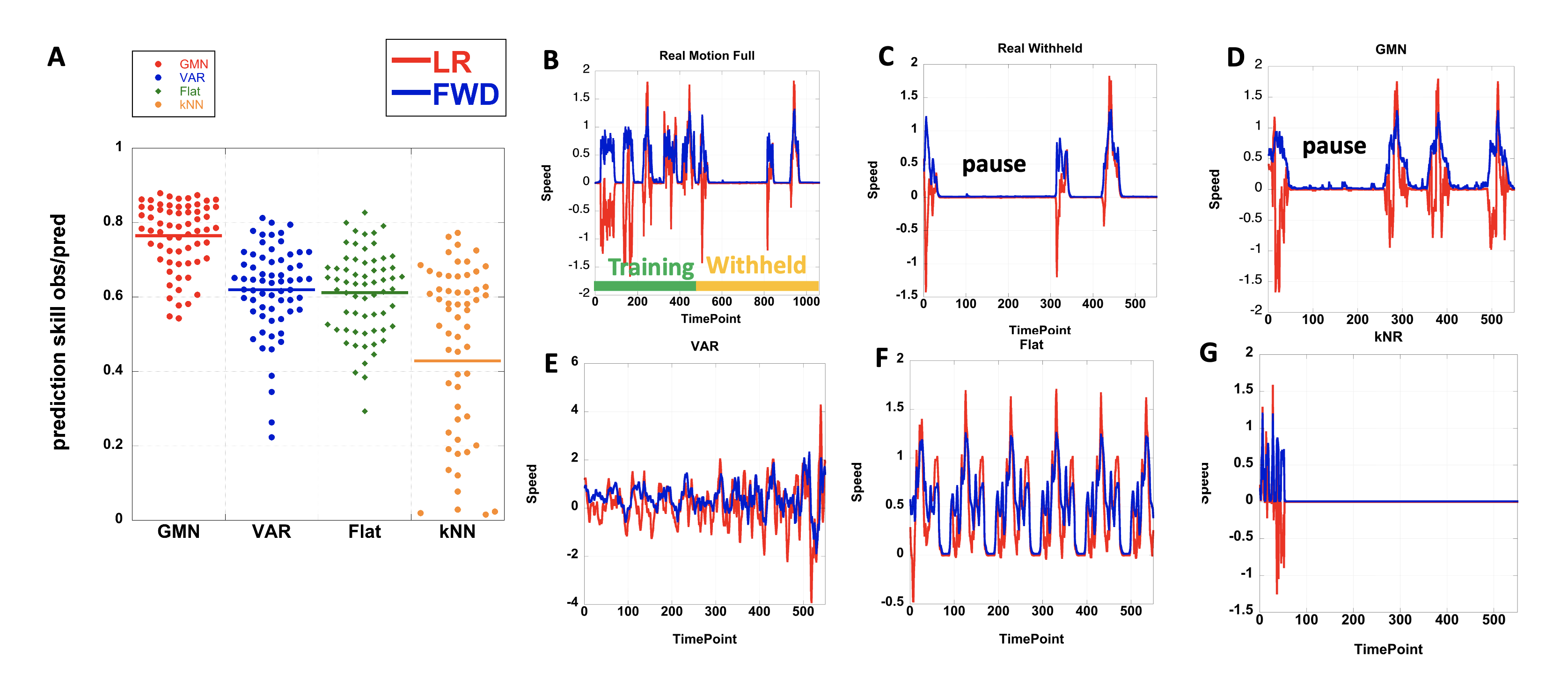}
  \caption{(A) Prediction skill of the 4 generative models on withheld data to assess realism of generated time series. Time series of 4 models used in this study (B-F) In the GMN observe the similarity in the pauses not present in the training set but present in the withheld real time series.}
\end{figure}

\begin{figure}
  \centering
  \includegraphics[width=1\columnwidth]{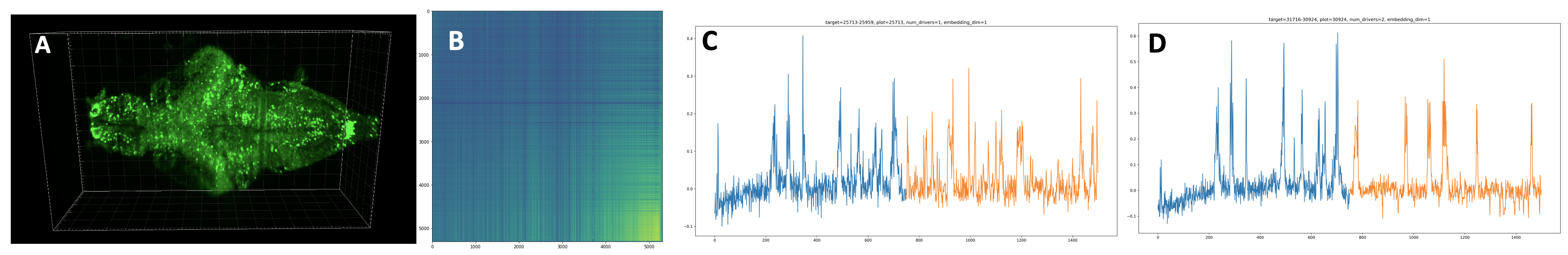}
  \caption{Scalability to Whole larval zebrafish brain GMN at single neuron resolution (A) Whole brain imaging of larval zebrafish with nuclear GCAMP6 (B) CCM Causal interaction matrix of >53K neurons(C, D) representative GMN generated (orange) left (C) and right(D) ventral spinal cord neuron dynamics from whole brain activity from whole brain activity (training period in blue).}
\end{figure}
\section{Discussion}
\label{discussion}
All optical electrophysiology has allowed to obtain neural activity datasets on whole brains from  experimental model organisms. This type of data when coupled with simultaneous quantitative behaviors and our manifold networks, allows us to uncover possible mapping functions between neural activity and behavior. Here, we have developed a method based on the generalized Takens theorem for uncovering such a mapping. 
This approach uses causal inference to reduce dimensionality and select the variables that are most informative and least redundant to identify candidate variables that can be used to build a network of low dimensional manifolds. Unlike other manifold representations that generate a single global representation of neural activity through dimensionality reduction that precludes experimental manipulation of the compressed dimension, our approach allows direct correspondence of the model to real observed features and hence it can be interrogated experimentally through electrophysiological, optogenetic or pharmacological interventions which can connect theory to empirical verification. 
Our network connects through shared nonredundant information which will have a topology that is very different from the network topology of the physical connectome (which is expected to be highly correlated in time). 
To use an analogy with a city, this is not a street map of the city but the relationships within the city parts extracted from the traffic patterns. 
Thus the causal network generated will capture relationships in the brain that reflect information flow but that will not have the same network structure as the physical brain. 
The nonredundancy aspect of our approach is important in order to properly ``unfold'' the manifolds so points will be optimally separated to avoid singularities and ambiguitiy in the predictions. 
In the current work, the $\rho_{\mbox{diff}}$ matrix was sufficiently sparse that redundancy was naturally avoided in the manifold generation scheme. In future, we envisage that an additional step will be necessary to avoid redundancy in the manifold variable selection, such as clustering the variables according to their causal relationships and enforcing the constraint that each variable in each manifold come from a different cluster. Preliminary results using a whole brain larval zebrafish dataset at single neuron resolution \cite{watanakeesuntorn2020massively} during a hypoxic escape response shows that ventral spinal cord motor neuron activity can be predicted using whole brain neuronal dynamics (Figure 8). Thus GMN scales to over >50,000 neurons and at single neuron resolution. Similarly fMRI datasets with human behavior also appear to have predictive power (supplementary materials).

Previous approaches using low dimensional manifolds to generate simulated behaviors (e.g. \cite{Sussillo2009}) have generated realistic human locomotion, fish swimming \cite{johnson2020probabilistic} based on kinematic data of motion capture, or neuronal dynamics using a variational autoencoder strategy \cite{pandarinath2018inferring}. However all these strategies do not allow experimental verification as latent variables do not have direct correspondence to real world neural structures, which GMN seeks to address.  
With reservoir computing and low dimensional embeddings it has been observed that when driven recursively, these models frequently fall into short repeating probably infinite cycles (e.g. \cite{Pathak2018}). With generative manifold networks, we seem to largely avoid locally ``trapped'' dynamics and manifold prediction collapses, especially in well connected nodes. 
The remarkable pauses that our network generated that were not in the training data (but that were in the testing data) are not at a nonzero value close to the median of the distribution, and they are of limited duration, after which the network reinitiates normal looking large amplitude oscillatory behaviors. 
We believe that it is possible that these could be an emergent property of hidden information contained within the network that was not explicitly present in the training set and is an interesting area for further investigation. 
The dataset we used to generate this model was essentially from a fly to which no stimulus was given and allowed to freely move that should be akin to a “default network”. In future work we wish to incorporate artificial sensory input and test in fMRI settings where both temporal and spatial resolution are limited (supplementary materials). 
The GMN technology can in principle reproduce any arbitrary behavior from recorded neural data provided it is sufficiently complete and contains information on the key variables relevant to the task. Insofar as this technology brings us closer to the far flung possibility of downloading a brain into a computer (a naturally seeded AI that can then generate natural behaviors), the consequences could be far reaching. The technology itself is pretty neutral though it could be conceivably misused for example in the violation of peoples privacy by capturing their thoughts if applied to imaging. New AI/ML algorithms can have societal impacts that are not estimable at this point. 
In addition to neuroscience, we believe that this type of network architecture is generic and applicable to any nonlinear network, so it may have wide ranging applications in other areas.

\clearpage
\newpage
\printbibliography
\newpage

\input{neurips_2021_supplementary}

\end{document}

%% file: neurips_2021_supplementary.tex
\setcounter{figure}{0}

\makeatletter 
\renewcommand{\thefigure}{S\@arabic\c@figure}
\makeatother

\section*{Supplementary data}

\begin{figure}[hbt!]
  \centering
  \includegraphics[width=1\columnwidth]{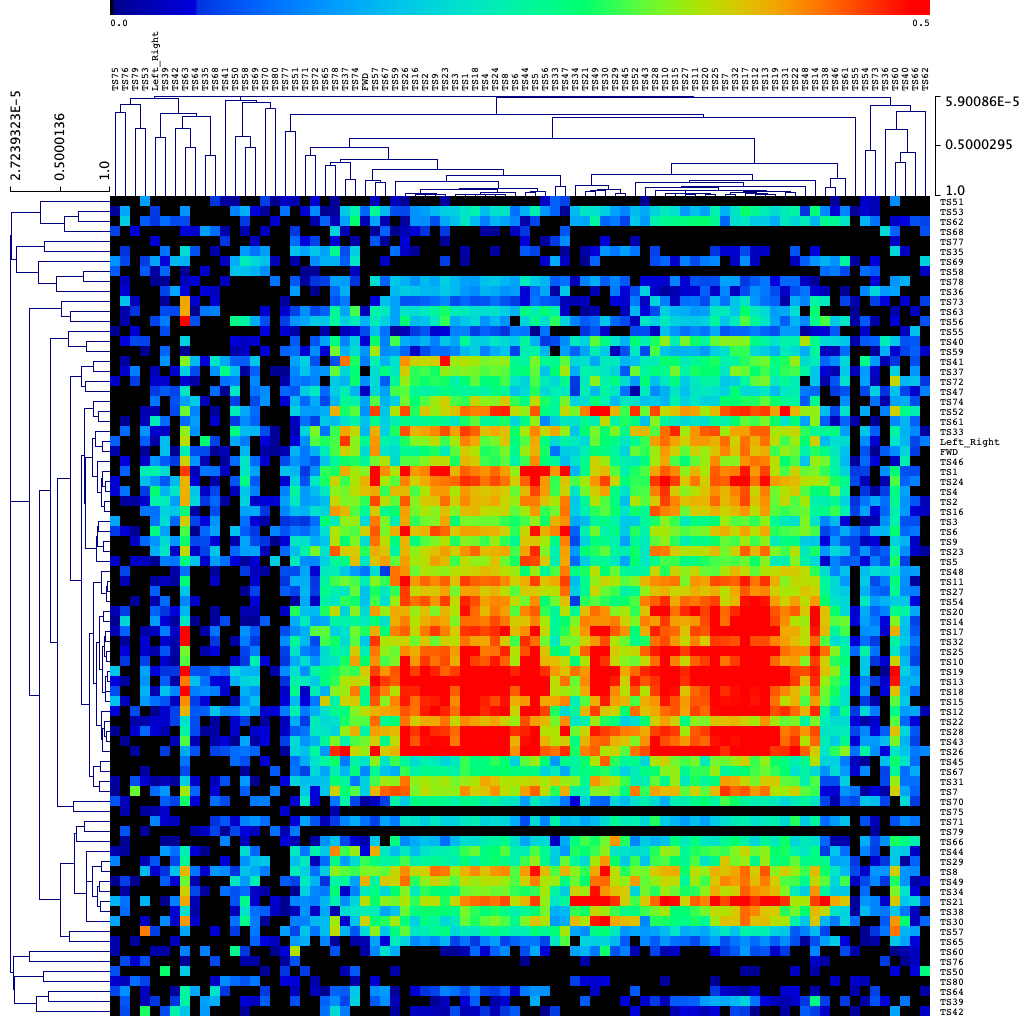}
  \caption{Convergent Crossmapping (CCM) causal strength Rho (observed/predicted Pearson correlation) matrix of 80 brain areas as described in (1) and Left/Right and forward (FWD) speed time series clustered by Pearson correlation for both inputs and outputs of the network. Note that CCM does not distinguish between direct and indirect causal interactions.}
\end{figure}

\newpage

\begin{figure}[hbt!]
  \centering
  \includegraphics[width=1\columnwidth]{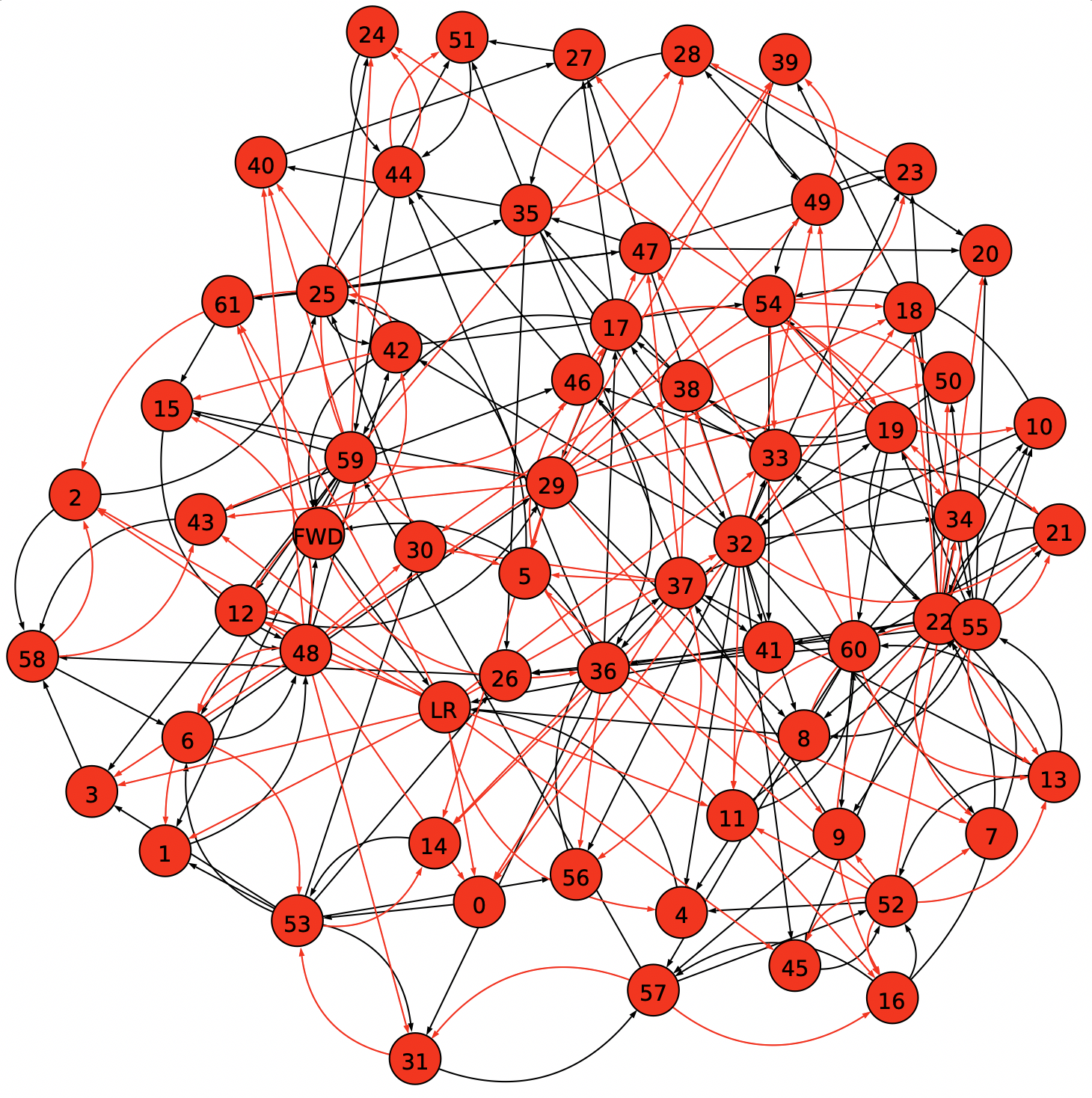}
  \caption{Network reconstructed for the Drosophila walking imaging dataset for the construction of the Generative Manifold Network. Black edges are primary drivers and Red edges follow to update the network state. The chosen embedding dimension for this network is 4.}
\end{figure}

\newpage

\begin{figure}[hbt!]
  \centering
  \includegraphics[width=1\columnwidth]{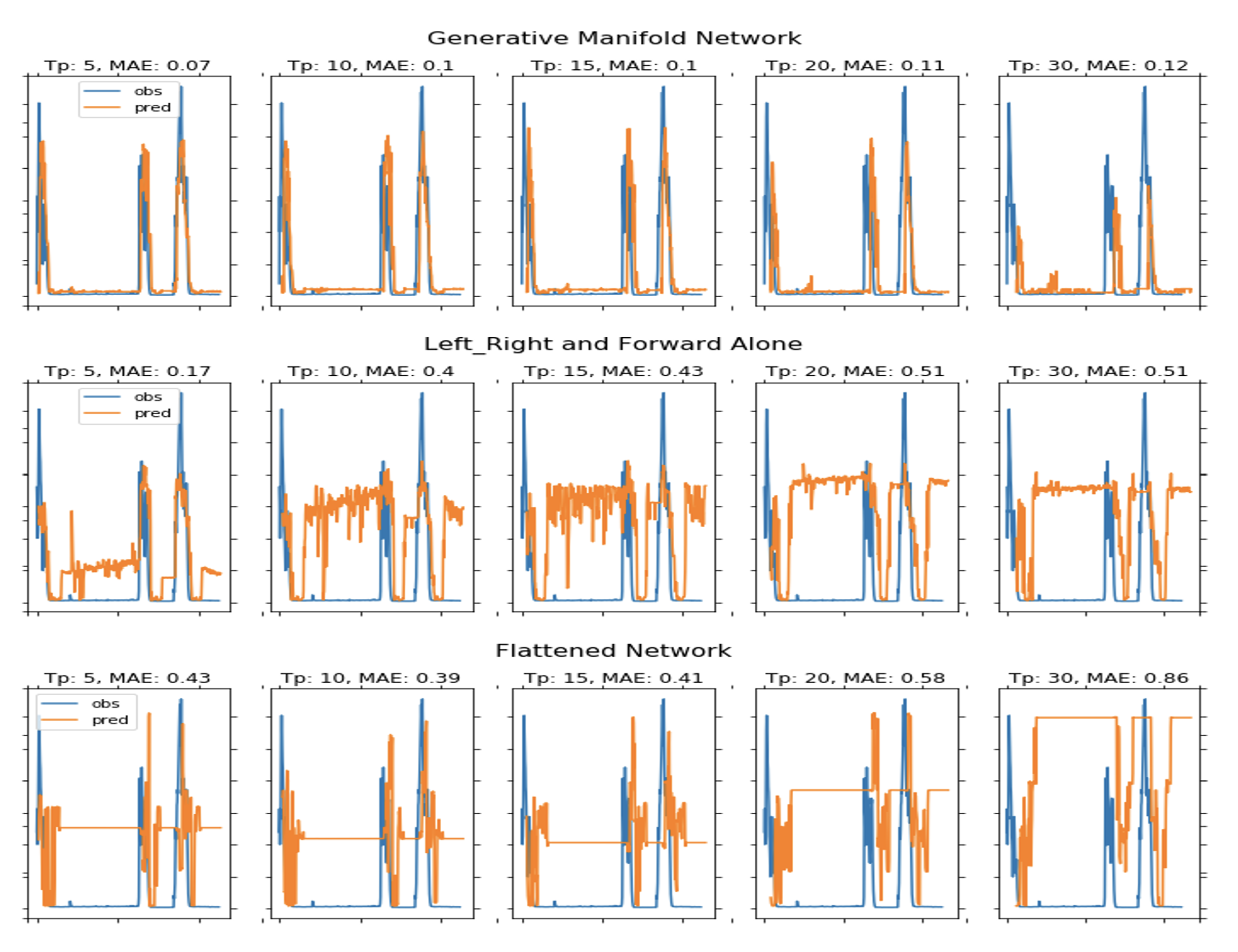}
  \caption{Out of sample validation. Using embeddings of the generated data to reconstruct attractors to predict withheld movement speed time series of the real fly. Workflow is as follows: 1. Use the generative manifold network, an embedding of the left/right and forward movement or a 82 dimensional manifold using all time series and the left right and forward movement speed time series to generate new time series through recursive predictions. Each method will predict a step into the future and use the prediction to update the model state recursively. Predicted time series are then used for an univariate embedding and used to predict withheld data. \\ 3 modes of prediction are shown with increasing prediction horizon Tp from 5 steps into the future to Tp = 30 time steps into the future. Real withheld time series are shown in blue, whereas predictions are shown in orange. Prediction error is given as mean absolute error (MAE)}
\end{figure}

\newpage

\begin{figure}[hbt!]
  \centering
  \includegraphics[width=1\columnwidth]{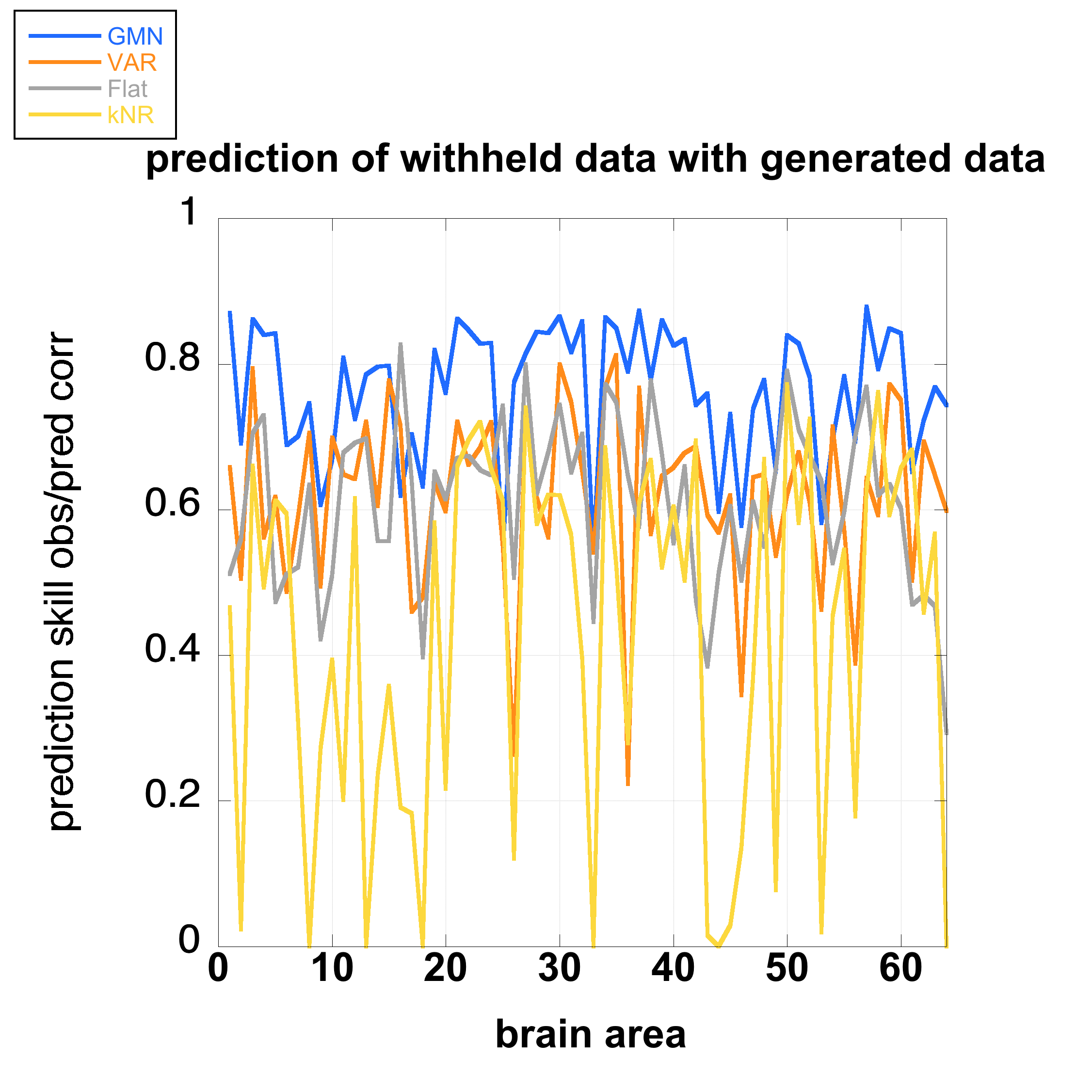}
  \caption{Prediction skill of the Generative Manifold Network (GMN) compared to a Vector Autoregressive model (VAR) a flat network (See figure S3 for description) and a k -Nearest Neighbor model. Models were used in generative mode to generate artificial time series. Time series are then univariately embedded to predict withheld real data for each brain activity time series as well as the forward speed and left right speeds using the simplex projection. Skill of the generated data embedding of each brain area are then tested for their ability to predict real withheld data not used to build the model. The ability of the synthetic data embedding to predict real withheld data is used as a metric of realism of the simulated, generated data. The skill of prediction is measured as the ability to predict real withheld data using generated artificial data from the GMN. Observe that in almost every case prediction accuracy of univariate embeddings using the GMN generated time series (see blue line generally above others)) outperform all other methods. Performance is measured as the mean observed/predicted Pearson correlation coefficient for all the points predicted across the entire withheld time series. 63 neuronal activity time series were used as well as 2 movement speed regressors (left/right speed and forward speed of the fly). Points between brain areas prediction performance of the various models are connected for visualization purposes to allow easier comparison of the models across brain areas. The connection does not imply any sequential relationship.}
\end{figure}

\newpage

\begin{figure}[hbt!]
  \centering
  \includegraphics[width=1\columnwidth]{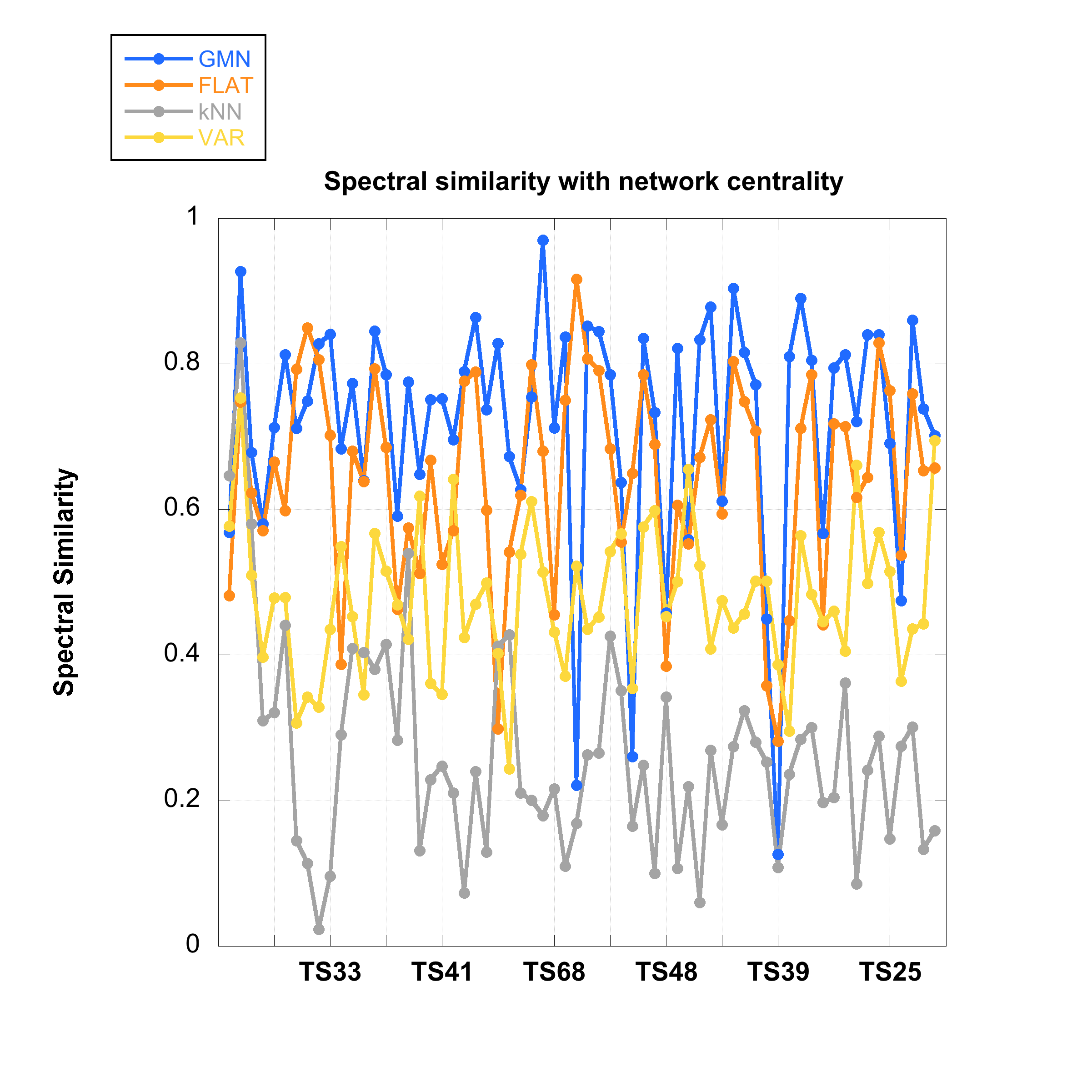}
  \caption{Prediction skill of the Generative Manifold Network (GMN) compared to a Vector Autoregressive model (VAR) a flat network as measured by spectral similarity.  (See figure S3 for description) and a k -Nearest Neighbor model. Models were used in generative mode to generate artificial time series. Fourier spectrum of the generated time series were then compared the spectrum of withheld real data for each brain activity region as the forward speed and left right speeds. Similarity score the Pearson correlation between the spectrum of the generated data and the spectrum of the withheld real time series.  The ability of the synthetic data embedding to produce similar spectra is used as a metric of realism of the simulated, generated data. In most cases spectral similarity of  the GMN generated time series (see blue line generally above others)) outperform all other methods. Points between brain areas prediction performance of the various models are connected for visualization purposes to allow easier comparison of the models across brain areas. The connection does not imply any sequential relationship. Spectral similarity of the time series were ordered according to network centrality rank. Results show a predominantly flat distribution suggesting that all parts of the network are more or less equally well described.}
\end{figure}

\newpage

\begin{multicols}{2}

\begingroup
  \centering
  \includegraphics[width=1\columnwidth]{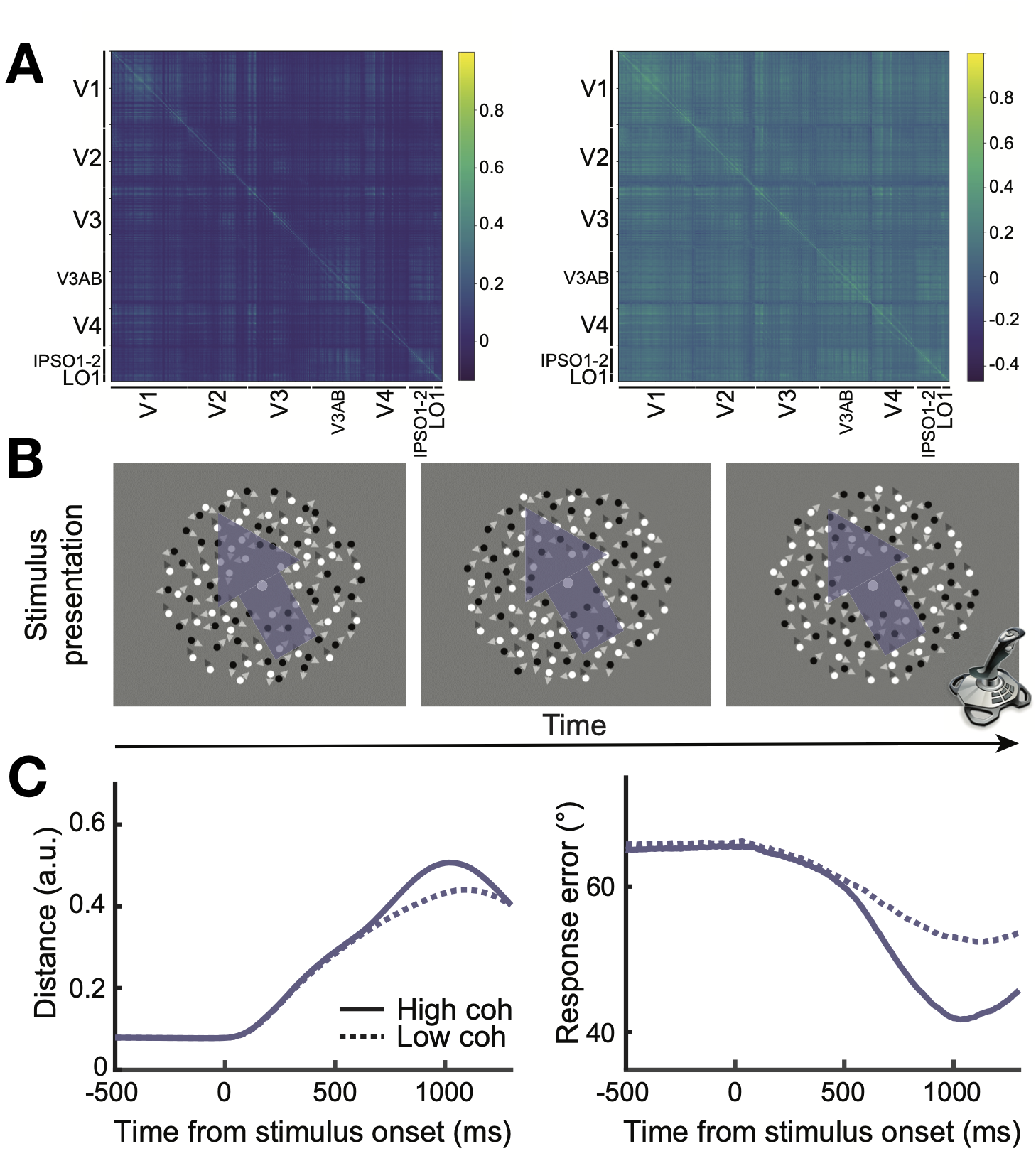}
  \includegraphics[width=1\columnwidth]{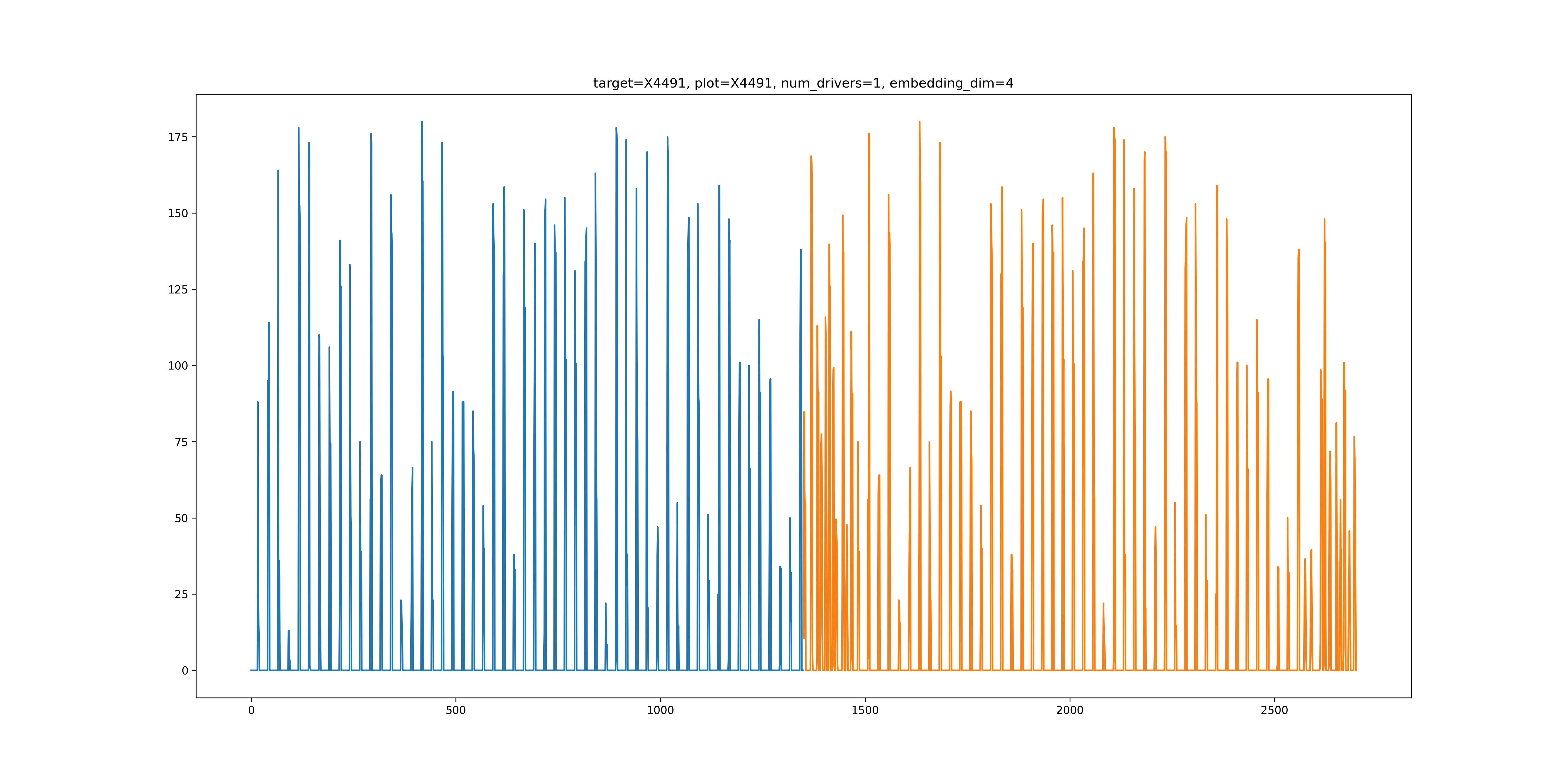}
  \includegraphics[width=1\columnwidth]{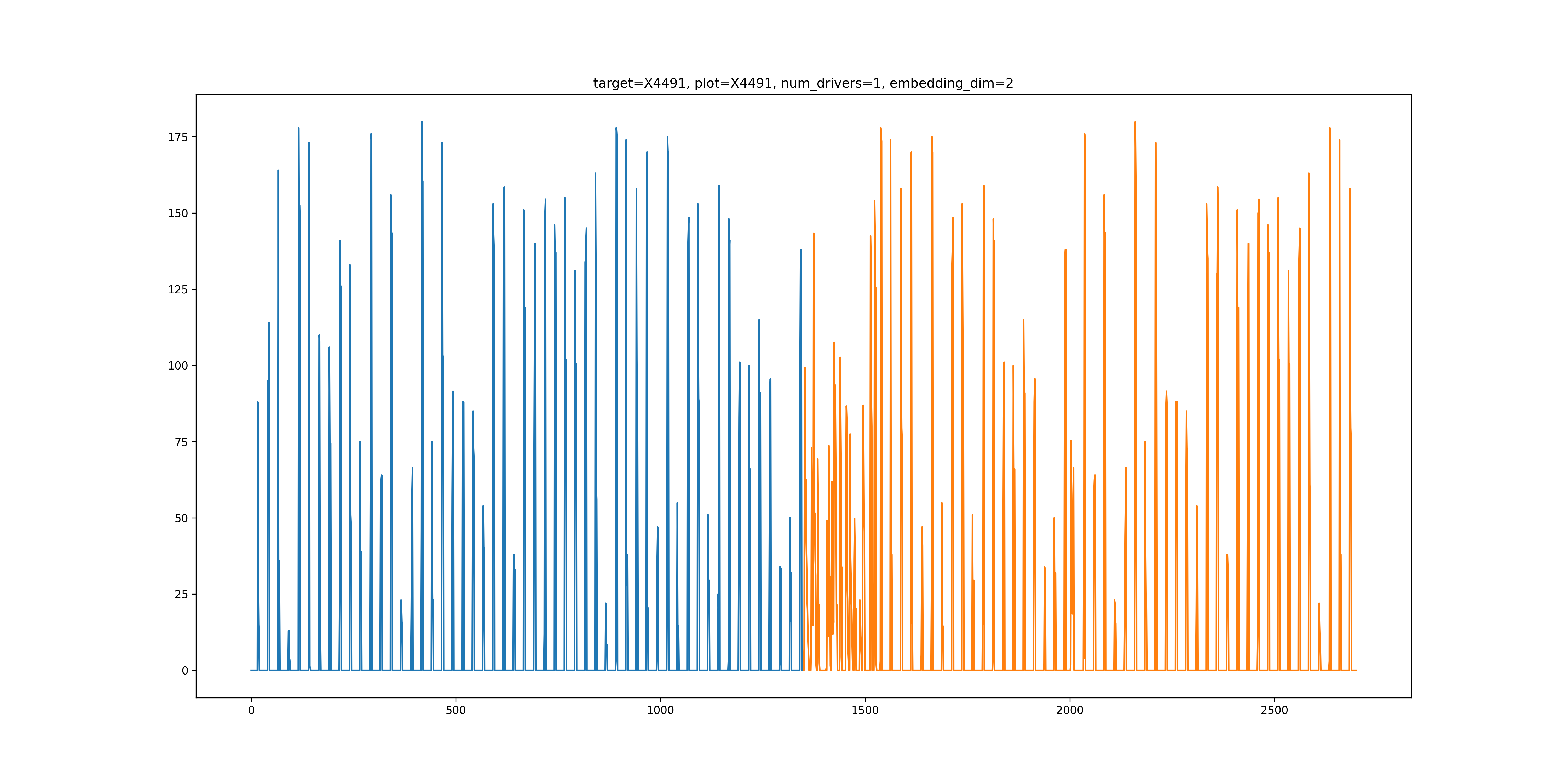}
 \captionof{figure}{A. (left) CCM causal network inference across 8 regions of interest from an existing fMRI dataset collected while participants performed a visual working memory task. Time series of 4492 voxels of the regions of interest used as well as the subject output (right) Simple correlation matrix associated with each region of interest. \\
  B. A sample trial of the fMRI experiment. On each trial, participants monitor a display of randomly moving dots for a series of brief target displays of coherent motion that is made up of either black or white dots. Participants report each of the coherent motion detected by continuously moving a flight simulator joystick along a trajectory (0\textdegree to 360\textdegree). \\
  C. Fine-grained behavioral dynamics can be extracted alongside neural activities recorded by fMRI through the measures of response trajectory and response error over time. Last 2 panels of Fig. S6C show two successful conditions where the GMN network generates realistic behavior of the fMRI and human behavioral output embedding. Training data is shown in blue for both cases  and generated data in orange. Generated simulated behaviors were generally in agreement although showed some short periods of higher frequency responses than in the training set. Examples shown differed in their embedding dimensionality, being five dimensional in the first example and three dimensional in the bottom example. Data was originally collected by Rademaker et al. (R. L. Rademaker, C. Chunharas, J. T. Serences, Coexisting representations of sensory and mnemonic information in human visual cortex. \textit{Nat. Neurosci.} 22, 1336–1344 (2019).)}
\endgroup

\end{multicols}

\newpage

\begin{figure}[hbt!]
  \centering
  \includegraphics[width=1\columnwidth]{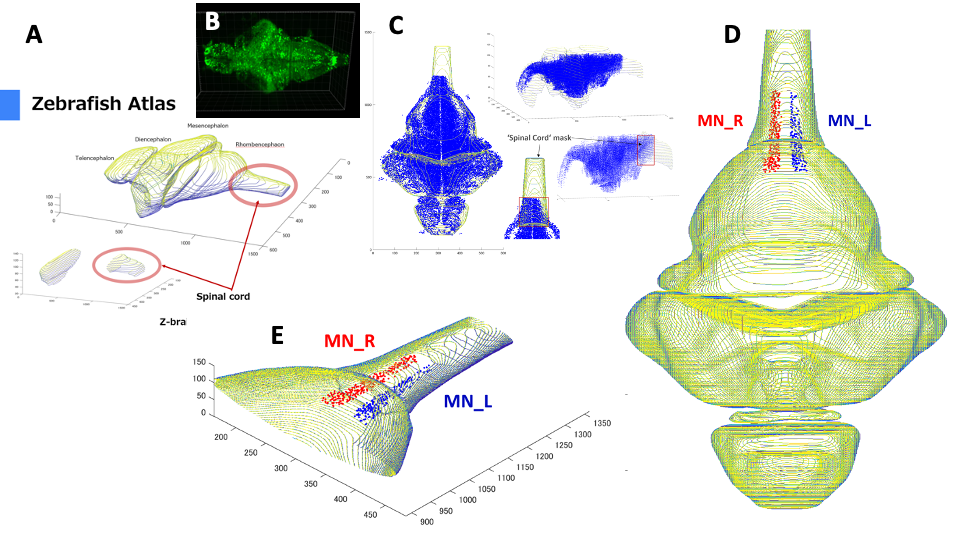}
  \caption{Building a realistic Generative Manifold Network whole brain simulation of the larval zebrafish at single neuron resolution. Transgenic larval Zebrafish expressing the GCAMP6f Calcium indicator localized to the nucleus of neurons was used for all optical electrophysiology measurements on a selective plane illumination microscope (B). As fish are completely embedded in agarose due to difficulty of image registration without immobilization movements could not be measured directly on the fish. As a surrogate for actual movement we selected both left and right motor neurons which are localized ventrally in the spinal cord. For this we registered our images with the Zebrafish atlas template (A, C). Using this map we localized both the left and right motor neuron tracts within the spinal cord (D,E). These were evaluated and the reconstructed GMN whole brain activity was used to drive the activity of these motor neurons}
\end{figure}

\newpage

\begin{figure}[hbt!]
  \centering
  \includegraphics[width=1\columnwidth]{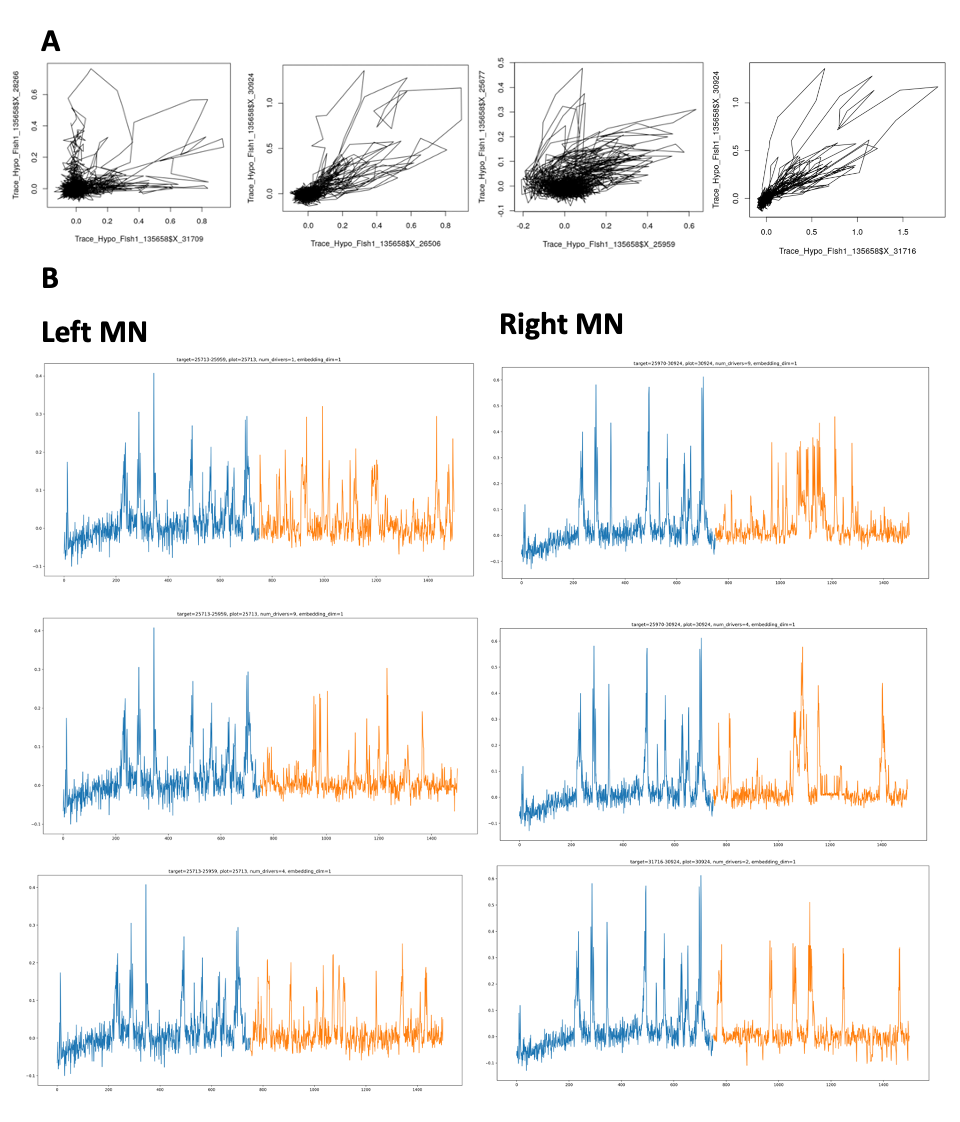}
  \caption{Building a realistic Generative Manifold Network whole brain simulation of the larval zebrafish at single neuron resolution. Pairs of putative left and right motor neurons activity phase portraits are consistent with the correct identification of motor neurons. Shown are 4 examples of left and right motor neuron pairs (A) whose activity is plotted against each other and show phase portraits consistent with what would be expected of left right coordination during swimming. . (B) Simulated motor neurons of the left and ventral spinal cord show realistic dynamical behaviors. Training data used is shown in blue and subsequently GMN generated artificial motor neuron behaviors are in orange. Shown are 3 representative pairs of GMN simulated left and right neurons showing behavior consistent with their real counterparts.
}
\end{figure}

\newpage

Hardware resources used were part of the ABCI supercomputer of the National Institute of Advanced Industrial Science and Technology in Kashiwanoha, Japan, For CCM calculations we used the mpEDM implementation optimized for GPU utilization and Open MPI parallelization. https://ieeexplore.ieee.org/document/9359204 and for network reconstruction and simulation using GMN the algorithm described in the present paper. GMN was run on the NAIST HPC cluster in a CPU only implementation. Description of the hardware configurations follow below. 

ABCI - Computing Node (V) - 512 nodes \\
Model: Fujitsu Primergy CX2570 \\
CPU: Intel Xeon Gold 6148 × 2 \\
GPU: NVIDIA Tesla V100 SXM2 (16GB HBM2) × 4 \\
Memory: 384GB DDR4 2666MHz \\
Local Storage: 1.6TB NVMe SSD (Intel SSD DC P4600 u.2) × 1 \\
Interconnect: InfiniBand EDR (100Gbps) × 2

NAIST HPC Server - Large Shared Memory Node - 2 nodes \\
Model: Oracle Server X5-8 \\ 
CPU: Xeon E7-8895v3 × 8 \\
GPU: None \\
Memory: 2TB \\
Local Storage: 1.2TB HDD × 6 \\
Interconnect: InfiniBand QDR (40Gbps) × 2

* Corresponding Author GMP geraldpao@gmail.com, g1pao@ucsd.edu, pao@salk.edu

Acknowledgements:
This work was funded by the Kavli Institute for Brain and Mind (KIBM, UCSD) with an Innovative Research Grant award to GMP.

COI Disclosure: GMP and CS are inventors on a patent filing claimed by the Salk Institute for the Generative Manifold Network algorithm.